# Some Spreadsheet Poka-Yoke

Bill Bekenn and Ray Hooper
Fairway Associates Ltd. PO Box 846, Ipswich IP5 3TT UK
info@fairwayassociates.co.uk

**ABSTRACT**

*Whilst not all spreadsheet defects are structural in nature, poor layout choices can compromise spreadsheet quality. These defects may be avoided at the development stage by some simple mistake prevention and detection devices. Poka-Yoke (Japanese for Mistake Proofing), which owes its genesis to the Toyota Production System (the standard for manufacturing excellence throughout the world) offers some principles that may be applied to reducing spreadsheet defects. In this paper we examine spreadsheet structure and how it can lead to defects and illustrate some basic spreadsheet Poka-Yokes to reduce them. These include guidelines on how to arrange areas of cells so that whole rows and columns can be inserted anywhere without causing errors, and rules for when to use relative and absolute references with respect to what type of area is being referred to.*

## 1    INTRODUCTION

The unstructured nature of a spreadsheet, allowing any layout and any type of formula, is a contributing factor to spreadsheet defects. There are no generally accepted good practice guidelines for spreadsheet development. Moreover, defects are often found long after the mistake has been made. Defect reduction is usually done by time consuming manual inspection (audit).

Poka-Yoke (pronounced "POH-kah YOH-kay") or mistake proofing, was first identified by Shigeo Shingo [Shingo, 1986] in the 1960s when, as a statistical process control engineer, he became frustrated that he could not achieve zero defects in manufacturing processes. Shingo realised that there was a clear distinction to be made between a mistake and a defect. A mistake was something that would inevitably lead to a defect unless one had a method to prevent or detect it within the manufacturing process. He also realised that "mistakes" were not always the fault of the operator particularly as the consequent defect becomes visible only at a later stage of the manufacturing process. The spreadsheet developer has little within Excel to help avoid defects and virtually nothing to aid good layout practice. In spreadsheets, defects can go undetected [EuSpRIG, 2007]. The goal of a Poka-Yoke device is to eliminate defects by preventing mistakes from occurring or at least detecting them at source rather than correct, at much greater cost, by audit when they become defects or errors (see Figure 1).

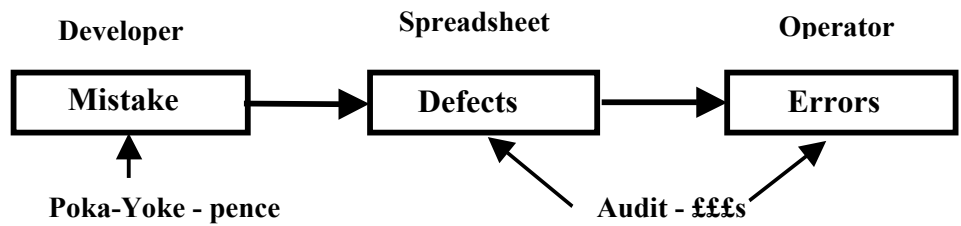

Figure 1. The Poka-Yoke Advantage

Well-known Poka-Yokes include the floppy-disk which can only be inserted one way into a drive. Software development, too, has its Quality Control methods [Rajalingham, 2000]. Indeed structured programming and object oriented design are, in a sense, prevention Poka-Yokes [Robinson, 1997].

In this paper we will describe defects due to poor layout and inadequate formula design and how they relate to Excel's vulnerabilities. Structures and practices that can be employed to minimise defects during spreadsheet development and modification are shown. This includes the rules required for layout and the correct use of relative, absolute and mixed references. These rules, supported by our *F*ormula*D*ata*S*leuth® ("Sleuth") application [Bekenn, 2007], help to provide structure to a spreadsheet.

Moreover, the Sleuth implements SPYs (Spreadsheet Poka-Yokes). Both prevention and detection SPYs are described along with how they help developers to reduce spreadsheet defects. Some of these SPYs, e.g. good sheet layout and including an additional blank row/column at the bottom/right of a range, are encouraged by the use of the Sleuth but can be implemented without it. Other SPYs, e.g. grouping related areas and replication, are available as part of the Sleuth's procedures.

### 1.1 Conventions

Some conventions used in this paper are:-

1. Data and formulas are not intermingled; they are always in separate blocks,
2. Spreadsheet structure and formula design is based upon Fill,
3. For the purposes of consistency and clarity we have adopted a convention that puts items in rows and instances in columns,
4. Data areas and cells are coloured Light Yellow in most examples,
5. Column letters and row numbers are shown top and left in most examples,
6. Some examples of referencing previous columns, inserting columns etc. are extendable to rows also.

### 1.2 Good Poka-Yokes (SPYs) Features

1. they are simple. If they are too complicated they will not be used,
2. they are part of the development process,
3. they are placed where the mistakes occur, to provide quick feedback.

## 2 STRUCTURAL DEFECTS

### 2.1 Mistakes

Human error has been studied in relation to spreadsheet mistakes [Panko, 2005]. However, errors can arise from defects particularly where modifications are being applied to existing workbooks. Such defects may arise from layout or formula mistakes. Some examples of mistakes and defects resulting in errors are listed below:-

1. Incorrect filling of formulas after modification due to the defect of the left column (or top row) formula being different to the rest.
2. Accidentally inserting rows/columns through unrelated areas of a spreadsheet not visible on screen due to defective layout,
3. Omitting to insert additional rows/columns in all "connected" areas throughout a workbook,

4. Failure to fill all areas of formulas after inserting rows/columns,

Care has to be taken when modifying workbooks and the flow of calculation needs to be understood. Irregular formulas and other defects have to be found so that filling from the top left and inserting rows/columns will not cause errors.

**2.2    Excel Vulnerabilities**

There are a number of types of error that occur due to Excel's in-built vulnerabilities, which themselves may be considered as defects. Some examples are listed below:-

1. Range references e.g. "SUM(C8:C9)", not adjusting when additional rows are inserted below the bottom of the range,
2. Formulas do not fill correctly due to incorrect use of relative, absolute and mixed references,
3. Incorrect formulas in a pasted area after copy and paste, due to clash of requirements for relative/absolute references needed for fill and copy.

**3    SPREADSHEET STRUCTURE**

Spreadsheet structure may be defined in terms of "Stripes", "Groups" of Blocks within Stripes, and "Replication", which are explained below.

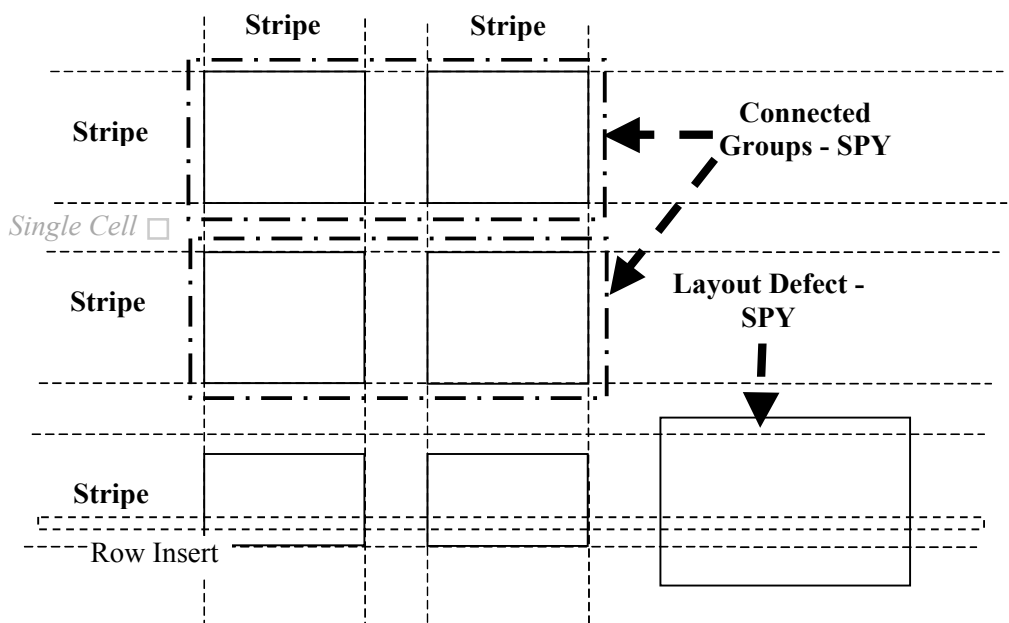

Figure 2 Spreadsheet Structure and some SPYs

It is also possible to conceive a Column or Row Insertion Test (CRIT) to test the resilience of a spreadsheet to modification. Any structure or formula which cannot have whole rows/columns inserted without causing errors is considered as having a CRIT defect.

**3.1    "Single Cells"**

"Single Cells" contain isolated items of data or one-off formulas and to fill requires an absolute reference, e.g. $C$12. Often Single Cells are constants used in a filled formula area. Constants written into filled formulas would be unseen, difficult to audit and a potential source of defects. A SPY could be devised to detect this mistake, but a SPY could be obtrusive because some constants may be permissible (e.g. 2, 10, 100, and 1000).

**3.2 "Blocks"**

An area of data or filled formulas is referred to as a "Block" [O'Beirne, 2005]. Blocks are considered distinct from Single Cells. There are three types of Block:-

1. Single Row: Columns can be inserted within the Block.
2. Single Column: Rows can be inserted within the Block
3. Multiple Row and Column: Columns and Rows can be inserted within the Block.

The absolute/relative row/column references (dollaring) that can be used in a filled formula must be chosen carefully (see section 10).

**3.3 "Stripe"**

A horizontal set of Blocks, spanning the same rows, make up a horizontal "Stripe". A vertical set of Blocks, spanning the same columns, make up a vertical "Stripe". Stripes are more important if they contain more than one row or more than one column.

**3.4 "Idealised Layout"**

An "Idealised Layout" is a sheet layout composed only of Stripes. These stripes can have whole rows or whole columns inserted at any point within a sheet, without introducing the risk of an error. Stripes are not allowed to overlap and must occupy their own set of rows/columns, but a row stripe can intersect a column stripe. Where possible, references within formulas are to cells on the same horizontal row or the same vertical column.

If selecting whole rows/columns slices through blocks of different sizes then it is likely that inserting rows/columns will result in a defect. The Sleuth contains a detection SPY for this so that the layout defect is avoided, see Figure 2.

**3.5 "Calculation Chain"**

A "Calculation Chain" starts from an input data item and finishes at a final result. Each non-data Block or Single Cell member of a Calculation Chain contains formulas that reference previous members of the Chain. Several Calculation Chains may exist in parallel giving different final results or they may combine at some point to form a single Calculation Chain.

**3.6 "Group"**

A horizontal or Row "Group" relates all the Blocks within horizontal Stripes of the same row height that are involved in a Calculation Chain. A vertical or Column "Group" relates all the Blocks within vertical Stripes of the same column width that are involved in a Calculation Chain. A Calculation Chain may flow through blocks of different heights/widths and thus several Groups can exist in the Chain. When rows/columns are to be inserted the insert must be repeated for all member Blocks of a Row/Column Group. The Sleuth contains a SPY to ensure that this happens by defining connected Groups, see figure 2.

 

# 4 CLASSES OF REFERENCE

Creation of defect-free formulas in a spreadsheet requires consideration of the different forms of reference.

## 4.1 Reference Type

The basic types of reference, "Relative", "Absolute" and "Mixed" are defined in [O'Beirne, 2005]. A relative reference locates another cell with respect to the position of the cell containing the formula (un-dollared),
	e.g. "=C3" in cell G5 translates to "=R[-2]C[-4]".
An absolute reference addresses a cell by its column and row address (dollared),
	e.g. "=$C$12" filled in cells G12 to I13 translates to "=R12C3".
A mixed reference is absolute row and relative column or relative row and absolute column.

## 4.2 Entities Referred To by Single or Range References

A spreadsheet formula is a combination of references and mathematical operators calculating a result. The references can be any one or a mix of the above types. Moreover, these can be either a Single Cell or a filled Block.. Figures (3) & (4) illustrate a range of possibilities.

For a single reference, e.g. =C3, = C7 or =$C$12:

|   | B | C | D | E | F | G | H | I | J | K | L | M | N | O | P |
|---|---|---|---|---|---|---|---|---|---|---|---|---|---|---|---|
| 2 | | | | | | | | | | | | | | | |
| 3 | | 10 | | | | =C3 | | | | | 0 | 20 | 65 | 85 | Cumulative |
| 4 | | 1) Single Cell to Single Cell | | | | | | | | | | =M3-L3 | =N3-M3 | =O3-N3 | Incremental |
| 5 | | | | | | | | | | | | 4) Previous Column Reference | | | |
| 6 | | | | | | | | | | | | | | | |
| 7 | | 5 | 30 | 25 | | =C7 | =D7 | =E7 | | | | | | | |
| 8 | | 20 | 65 | 85 | | =C8 | =D8 | =E8 | | | | 20 | 45 | 20 | Incremental |
| 9 | | 2) Block to Block | | | | | | | | | 0 | =M8+L9 | =N8+M9 | =O8+N9 | Cumulative |
| 10 | | | | | | | | | | | | 5) Self Reference to Previous Column | | | |
| 11 | | | | | | | | | | | | | | | |
| 12 | | 45 | | | | =$C$12 | =$C$12 | =$C$12 | | | | | | | |
| 13 | | | | | | =$C$12 | =$C$12 | =$C$12 | | | | 20 | 65 | 85 | =O13 |
| 14 | | 3) Block to Single Cell | | | | | | | | | | 6) Partial Reference to a Block | | | |

Figure 3 Single References

The above situations are in order of potential ascending defect probability.

1. Any dollaring combination will work.
2. Reference must be relative (the default),
3. Reference must be absolute, (not the default), otherwise there is a defect leading to errors in the filled cells,
4. The formula has a CRIT defect because inserting columns will leave the formula to the right of the insert pointing to the wrong column; filling into the inserted blank cells only still leaves an error,
5. as for 4,
6. There is a CRIT defect, as inserting columns to the right of the block leaves the last column partial reference pointing to the wrong column and it may be unseen on another sheet.

For range reference =C3:D4 or = $C$3:$D$4:

```
     B    C         D         E            F G H I  J       K    L   M    N  O P Q          R
2
3         5         30                            5   30  25    45  70 55   =SUM(I3:O3)
4         20        65        =SUM(C3:D4)         20  65  85    65  85 35   =SUM(I4:O4)
5         1) Single Cell to Block                 3) Multiple Block Reference
6
7
8         5         30        =SUM(C8:D8)      0  5   30  25              =AVERAGE(H8:I8)   =AVERAGE(I8:J8)
9         20        65        =SUM(C9:D9)      0  20  65  85              =AVERAGE(H9:I9)   =AVERAGE(I9:J9)
10        =SUM(C8:C9) =SUM(D8:D9)                 4) Partial Reference to a Block
11        2) Block to Block
12
13                                                5   30  25  45  70  55  =AVERAGE(J13:L13) =AVERAGE(K13:M13)
14        20        65        Cumulative          20  65  85  65  85  35  =AVERAGE(J14:L14) =AVERAGE(K14:M14)
15   0    =C14-SUM($B15:B15) =D14-SUM($B15:C15) Incremenal
          3) Self Reference (also Complex Reference)  4) Partial Reference to Multiple Blocks
```

Figure 4 Range References

The above situations are in order of potential ascending defect probability. The blank rows/columns needed to avoid an obvious CRIT defect are omitted to simplify figure 4.

1. Any dollaring combination will work,
2. Reference must be relative column in C10:D10 and relative row in E8:E9,
3. Left half of complex reference must be absolute column (not the default),
4. The gap between the blocks is a defect as items placed in the gap, e.g. a temporary SUM(), cause an error as they are included in the column Q SUM(),
5. Some data may not be accessed anywhere due to the partial nature of the reference, the formula has a CRIT defect as inserting columns will cause errors,
6. as for 5 and the blocks must remain adjacent as the formula has a CRIT defect.

### 4.3 "Complex Reference"

A "Complex Reference" is a range reference where the start cell to the left of the colon is a different type of reference to the end cell to the right of the colon (see figure 5). When filled, the width or height of the range will be different in each cell. It can be used, for example, instead of previous column and self references for accumulating, thus removing the need for a Seed Value and still avoid a CRIT defect. Complex references are described in more detail in the Appendix (see section 10).

```
    B  C    D        E        F        G           H I J   K          L           M
2
3        20       45       20       Incrmental     20      45         20          Incrmental
4   0    =D3+C4   =E3+D4   =F3+E4   Cumulative     =SUM($J3:J3) =SUM($J3:K3) =SUM($J3:L3) Cumulative
5        1) Cumulation with Seed Value             2) Cumulation with Complex Reference
6
```

Figure 5 Complex References

### 5 GOOD LAYOUT AND FORMULA QUALITY

Spreadsheets are different to paper-based sets of calculations (e.g. accounts or journals). Spreadsheets can be easily modified whereas paper-based accounts or journals cannot be modified easily. However, it may be necessary to have a spreadsheet output presented in a similar form to accounts or journals but this does not mean underlying calculation process has to be laid out that way. A good spreadsheet layout is a) easy to understand and b) easy to modify. Under modification each action may have consequential actions, e.g. inserting rows requires re-filling of formulas. Ease of modification can be assessed by achieving a low number of consequential actions following each original action. Good Layout and Formula Quality are very closely related.

|   | B | C | D | E | F | G | H | I | J | K | L | M | N | O | P |   |
|---|---|---|---|---|---|---|---|---|---|---|---|---|---|---|---|---|
| 2 |   |   |   |   |   | Quarters "Group", "Stripe" 1 | | | | | | Quarters "Group", "Stripe" 2 | | | | |
| 3 |   |   |   |   |   | Q1 | Q2 | Q3 | Q4 |   |   | Q1 | Q2 | Q3 | Q4 |   |
| 4 | Total Costs Inc. VAT | | | | | 164.5 | 270.25 | 352.5 | 521.7 |   |   |   |   |   |   |   |
| 5 |   |   |   |   |   | Incremental Quantities | | | |   |   | Cumulative Quantities | | | |   |
| 6 | Item 1 | | | | | 20 | 35 | 42 | 57 |   | 0 | 20 | 55 | 97 | 154 | Items "Group", |
| 7 | Item 2 | | | | | 10 | 15 | 22 | 36 |   | 0 | 10 | 25 | 47 | 83 | 2 row "Stripe" 1 |
| 8 |   |   |   |   |   |   |   |   |   |   |   |   |   |   |   |   |
| 9 |   |   | VAT | 17.5% |   | 17.5% | 17.5% | 17.5% | 17.5% |   |   |   |   |   |   |   |
| 10 |   |   |   | Cost |   | Quantities x Unit Costs Inc. VAT | | | |   |   | Percentage Increase | | | |   |
| 11 | Item 1 | | | 4 | | 94.00 | 164.50 | 197.40 | 267.90 |   |   | #N/A | 175% | 76% | 59% | Items "Group", |
| 12 | Item 2 | | | 6 | | 70.50 | 105.75 | 155.10 | 253.80 |   |   | #N/A | 150% | 88% | 77% | 2 row "Stripe" 2 |
| 13 |   |   |   |   |   |   |   |   |   |   |   |   |   |   |   |   |
| 14 | Total Costs Inc. VAT | | | | | 164.5 | 270.25 | 352.5 | 521.7 |   |   |   |   |   |   |   |

|   | E | F | G | H | I | J |
|---|---|---|---|---|---|---|
| 3 |   |   | Q1 | Q2 | Q3 | Q4 |
| 4 |   |   | =G14 | =H14 | =I14 | =J14 |
| 5 |   |   | Incremental Quantities | | | |
| 6 |   |   | 20 | 35 | 42 | 57 |
| 7 |   |   | 10 | 15 | 22 | 36 |
| 8 |   |   |   |   |   |   |
| 9 | 0.175 |   | =$E9 | =$E9 | =$E9 | =$E9 |
| 10 | Cost |   | Quantities x Unit Costs Inc. VAT | | | |
| 11 | 4 |   | =$E11*G6*(1+G$9) | =$E11*H6*(1+H$9) | =$E11*I6*(1+I$9) | =$E11*J6*(1+J$9) |
| 12 | 6 |   | =$E12*G7*(1+G$9) | =$E12*H7*(1+H$9) | =$E12*I7*(1+I$9) | =$E12*J7*(1+J$9) |
| 13 |   |   |   |   |   |   |
| 14 |   |   | =SUM(G11:G13) | =SUM(H11:H13) | =SUM(I11:I13) | =SUM(J11:J13) |

|   | L | M | N | O | P |
|---|---|---|---|---|---|
| 3 |   | Q1 | Q2 | Q3 | Q4 |
| 4 |   |   |   |   |   |
| 5 |   | Cumulative Quantities | | | |
| 6 | =0 | =L6+G6 | =M6+H6 | =N6+I6 | =O6+J6 |
| 7 | =0 | =L7+G7 | =M7+H7 | =N7+I7 | =O7+J7 |
| 8 |   |   |   |   |   |
| 9 |   |   |   |   |   |
| 10 |   | Percentage Increase | | | |
| 11 | =IF(L6=0,#N/A,(M6-L6)/L6) | =IF(M6=0,#N/A,(N6-M6)/M6) | =IF(N6=0,#N/A,(O6-N6)/N6) | =IF(O6=0,#N/A,(P6-O6)/O6) |
| 12 | =IF(L7=0,#N/A,(M7-L7)/L7) | =IF(M7=0,#N/A,(N7-M7)/M7) | =IF(N7=0,#N/A,(O7-N7)/N7) | =IF(O7=0,#N/A,(P7-O7)/O7) |
| 13 |   |   |   |   |   |

Figure 6. Example of layout and formulas for Sections 5.1 to 5.5

### 5.1 Using Seed Values to ensure Blocks fill.

Consider a spreadsheet consisting of four quarters laid out in two vertical stripes (G:J & M:P) as shown in figure 6. The first block (G6:J7) contains the quantities in each quarter and the second block (M6:P7) comprises formulas for calculating the cumulative numbers from the first stripe. If all Single Row or Multiple Row and Column areas, in the second block, are to fill from the left hand column the formulas cannot have exceptions. However, the first column is "different" because, for it to fill, it needs to refer to the previous column (L). This is achieved by ensuring the previous column is available for a "Seed Value" set to "=0" (column L). A good layout practice is therefore to reserve an additional column to the left of a block (F & L in this case) for seed values.

An alternative in this case is to use a Complex Reference which does not require a Seed Value, in M6, "=SUM($G6:G6)" filled to P6. A complex reference is more difficult to create correctly, but less defect prone.

**5.2  Stripe Based Idealised Layout**

Figure 6 shows a Stripe based layout which is aimed at making insertion of rows and columns easy and error free (addressing the second mistake identified in section 2.1). As Stripes do not overlap, whole rows or columns can be inserted at any point without damaging the structure. A simple prevention SPY in the Sleuth using the Automatic Grouping feature allows this to happen. Also in an idealised layout references in any formula should be vertical or horizontal to the same column or row that contains the formula. Although the Percentage Increase formula in figure 6 shows an exception to this, in general, formulas which reference vertically or horizontally are easier to create and understand and require less use of absolute referencing (dollar characters).

**5.3  Groups and Calculation Flow**

Calculation within a spreadsheet often flows though a number of blocks of the same size progressively refining the calculations towards a final result. It is often the case that, additional rows (or columns) are required in all the Blocks within the calculation flow.

Excel has no direct procedure for ensuring that all the Blocks have the additional rows (or columns) entered. The developer has either to remember where all the Blocks in a calculation flow are or use some form of spreadsheet coding/annotation to remind them. This can lead to a spreadsheet defect from the mistakes 3 & 4 in section 2.1. The Sleuth application, allows the definition of a Group which comprises all the Blocks which are connected (see figure 2). This enables the possibility of inserting a row (or column) in one Block with the automatic insertion of a row (or column) in all the connected Blocks. Moreover, the Sleuth re-fills the formulas in all member Blocks of the Group. This prevention SPY (mentioned in figure 2) requires only that the user defines the connected Blocks using the same Group Name.

Whilst the traditional flow of calculation in a spreadsheet tends to follow the paper based equivalents, i.e. top left to bottom right, this has disadvantages when viewing part of a sheet on a screen rather than a complete two page ledger. It is better to keep the important Blocks which are to be viewed most often on the left hand side. The flow can then proceed from top left to the far right and then work through intermediate calculations from right to left finishing with the results to be viewed in the left hand columns. The final result is likely to end up at the bottom left of a sheet and can be linked back to the top left where it is the first thing to be viewed (as in figure 6). Also this final result at the top left is unlikely to move as a result of column or row insertion and is thus a safer item to link to another workbook.

**5.4  Previous Column and Self References**

The Percentage Increase formulas in figure 6 show a previous column reference to subtract the previous quarter from the Cumulative Quantity. This involves the Seed Value in column L. The Cumulative Quantity formulas again refer to the previous column but this is also a Self Reference as apart from column Q, which refers to the seed value, the other columns refer to the previous column within their own Block. Both Previous Column and Self References are higher risk as they do not take values from the whole Block being referenced. This leaves a chance that the last column of the block may not be accessed anywhere and may be defective, e.g. the last quarter may not carry forward to the first quarter of the next period. Also previous column references are damaged by inserting columns, a CRIT defect, and must be re-filled across the whole block to avoid some cases leading to errors.

### 5.5 Partial Access

If another year were added to figure 6, (see Figure 7), then the seed values for the cumulative quantities in column L would become a Partial Access to cells P6 and P7 to carry forward quarter 4. Such references are high risk due to a CRIT defect as they do not access the whole block and insertion of columns (say to convert figure 6 from Quarters to Months) may leave them in error pointing to the wrong column.

|    |        |    |    |    | Q5 | Q6 | Q7 | Q8 |    | Q5 | Q6 | Q7 | Q8 |
|----|--------|----|----|----|----|----|----|----|----|-----|-----|-----|-----|
| 16 |        |    |    |    |    |    |    |    |    |     |     |     |     |
| 17 |        |    |    |    | Incremental Quantities | | | | | Cumulative Quantities | | | |
| 18 | Item 1 |    |    |    | 68 | 74 | 81 | 90 | 154 | 222 | 296 | 377 | 467 |
| 19 | Item 2 |    |    |    | 42 | 48 | 53 | 61 | 83  | 125 | 173 | 226 | 287 |
| 20 |        |    |    |    |    |    |    |    |    |     |     |     |     |

|    | G  | H  | I  | J  | K | L     | M         | N         | O         | P         |
|----|----|----|----|----|---|-------|-----------|-----------|-----------|-----------|
| 16 | Q5 | Q6 | Q7 | Q8 |   |       | Q5        | Q6        | Q7        | Q8        |
| 17 | Incremental Quantities | | | | | | Cumulative Quantities | | | |
| 18 | 68 | 74 | 81 | 90 |   | =P6   | =L18+G18  | =M18+H18  | =N18+I18  | =O18+J18  |
| 19 | 42 | 48 | 53 | 61 |   | =P7   | =L19+G19  | =M19+H19  | =N19+I19  | =O19+J19  |
| 20 |    |    |    |    |   |       |           |           |           |           |

Figure 7. Partial Access for Carry Forward

There are other cases where Blocks are referenced by a Partial Access, such as a rolling average over 3 months, and all are higher risk as some cells may end up not accessed and so not contributing to the final result. Also this type of partial access requires careful re-filling if columns are inserted as the formulas contain CRIT defects.

## 6 REPLICATION

Having created a set of calculations in a spreadsheet there is often a need to work on a second set of data, e.g. creating year 2 of a cash flow sheet from year 1. Copying the whole sheet to a new sheet is probably the safest way of achieving such a replication. If this second set is required below the original area, then Cut and Paste will move them from the temporary copied sheet. However, it is more convenient to Copy and Paste within a sheet, particularly when smaller areas of formulas are to be re-used. Unfortunately the formulas contain defects in how they will copy and paste but changing the dollaring would then cause defects with respect to fill. Copy and Paste or fill may convert these defects into spreadsheet errors. In fact, after pasting the new copy all the formulas require checking to ensure that combinations of relative and absolute referencing have produced the desired result in the pasted copy.

### 6.1 Apparent Adjustments Made by Copy and Paste

|    | B      | C | D | E   | F | G          | H       | I       | J      |
|----|--------|---|---|-----|---|------------|---------|---------|--------|
| 2  |        |   |   |     |   | Quantities |         |         |        |
| 3  | Item 1 |   |   |     |   | 20         | 35      | 42      | 57     |
| 4  | Item 2 |   |   |     |   | 10         | 15      | 22      | 36     |
| 5  |        |   |   |     |   |            |         |         |        |
| 6  |        |   |   |     |   |            |         |         |        |
| 7  |        |   |   |     |   | Quantities Plus Percentage 1 | | | |
| 8  | Item 1 |   |   | 20% |   | 24.00      | 42.00   | 50.40   | 68.40  |
| 9  | Item 2 |   |   |     |   | 12.00      | 18.00   | 26.40   | 43.20  |
| 10 |        |   |   |     |   |            |         |         |        |
| 11 |        |   |   |     |   | Quantities Plus Percentage 2 | | | |
| 12 | Item 1 |   |   | 10% |   | #VALUE!    | 0.00    | 0.00    | 0.00   |
| 13 | Item 2 |   |   |     |   | 28.80      | 50.40   | 60.48   | 82.08  |
| 14 |        |   |   |     |   |            |         |         |        |

|    | E   | F | G             | H             | I             | J             |
|----|-----|---|---------------|---------------|---------------|---------------|
| 2  |     |   | Quantities    |               |               |               |
| 3  |     |   | 20            | 35            | 42            | 57            |
| 4  |     |   | 10            | 15            | 22            | 36            |
| 5  |     |   |               |               |               |               |
| 6  |     |   |               |               |               |               |
| 7  |     |   | Quantities Plus Percentage 1 | | | |
| 8  | 0.2 |   | =G3*(1+$E$8)  | =H3*(1+$E$8)  | =I3*(1+$E$8)  | =J3*(1+$E$8)  |
| 9  |     |   | =G4*(1+$E$8)  | =H4*(1+$E$8)  | =I4*(1+$E$8)  | =J4*(1+$E$8)  |
| 10 |     |   |               |               |               |               |
| 11 |     |   | Quantities Plus Percentage 2 | | | |
| 12 | 0.1 |   | =G7*(1+$E$8)  | =H7*(1+$E$8)  | =I7*(1+$E$8)  | =J7*(1+$E$8)  |
| 13 |     |   | =G8*(1+$E$8)  | =H8*(1+$E$8)  | =I8*(1+$E$8)  | =J8*(1+$E$8)  |
| 14 |     |   |               |               |               |               |

Figure 8. Typical Copy and Paste Problems

Copy and Paste does not adjust formulas (unless pasting to another workbook). It simply copies the R1C1 formula strings to a new set of cells. In many cases such as Blocks referencing Single Cell inputs the references will be absolute so that they fill correctly.

The pasted formulas will thus refer to the original copy of the Single Cell inputs rather than to the new pasted copies. The converse will also happen where a Block refers to a common Block of inputs which is not being duplicated, these relative references will end up pointing to a relative position on the sheet which has no relation to the inputs required. Both these possibilities are shown in figure 8, where the original formulas from G8:J9 have been copied and pasted to G12:J13 and a new percentage value entered in E12. The right hand section of Figure 8 shows that the formulas have not translated in the way the user must have intended. The user is now committed to a number of consequent actions which heighten the possibility of a mistake and a resulting error.

**6.2    Replication as Intelligent Copy and Paste**

To avoid these defects the Sleuth has a Replication procedure which allows the user to progressively mark all the Blocks and Single Cells which are to be duplicated, and then replicate all of them in one shot. The duplicated copies of the Blocks and Single Cells are placed in an unused area below the area involved in the replication and Cut and Paste can be used to move them to wherever they are required. In this way the number of consequent actions is reduced and possible errors from this cause are removed.

References are adjusted following two rules; (i) a reference to a cell being replicated points to the new copy of that cell and (ii) a reference to a cell not being replicated continues to point to it. These adjustments are done irrespective of the relative/absolute reference types. This requires that the dollaring has to be correct in the original so that Blocks are filled correctly. But this is no more than would have been required for the original to work correctly without replication.

Replication is a form of intelligent copy and paste working to reduce the number of consequent actions required by the user when they need to duplicate areas. This feature acts as prevention SPY device.

**7    SPREADSHEET POKA YOKE**

The Sleuth's SPYs are summarised below together with the Mistakes, Defects and Errors that they address:-

1. Automatic fill of a Formula Block; prevents the use of different formulas in the left column / top row and encourages the use of Seed Values, ensures correct re-filling of blocks after changes to the formula,
2. Layout Defect Detection; warns when unrelated blocks of a different size are present in a Stripe and will prevent insertion of rows/columns if different groups overlap within a Stripe,
3. Connected Groups; removes the need for a developer to remember all the areas that interact and allows a one shot insert into all members of a Group with automatic re-filling of formula blocks,
4. Automatic re-filling of blocks when formulas are modified or rows/columns are inserted protects against errors caused by mistakes in re-filling or CRIT defects in formulas.
5. Automatic insertion of a blank row/column when inserting ensures that range references do not miss the additional rows/columns,
6. Precedents Reconciliation; detects Partial Accesses, mistakes in dollaring and references to cells outside of other blocks,
7. Formula Correction; adjusts the dollaring and the size of ranges to match the blocks being referenced,

8. Replication; ensures that references in the replicated areas are correct and removes the need for the checking of formulas that is needed after copy and paste,

## 8   CONCLUSION

This paper has examined spreadsheet structure and how "mistakes" can make a spreadsheet defective. These "mistakes" are sometimes due to Excel vulnerabilities not being appreciated, sometimes due to lack of forethought or awareness of how the spreadsheet might need to be modified. Drawing a parallel with the overwhelming success of notions of quality to reduce defects in manufacturing using the pioneering Toyota Production System we have suggested that Poka Yoke, i.e. mistake proofing, could be a valuable technique to reduce defects and therefore the risk of error.

Our detailed consideration of spreadsheet structure has lead to identifying a number of defects and some Spreadsheet Poka Yoke (SPYs) to prevent or detect the defects before they are propagated. Defect reduction at the development stage will potentially save time and effort on auditing. The application of a Column Row Insertion Test (CRIT) and the attendant SPYs in our *F*ormula*D*ata*S*leuth® (Sleuth) application along with the Replication SPY has been tested in a real application for a complex cost-modelling scenario, which will be reported.

Although we have not considered the out-and-out human error aspect of "mistakes" it is our belief that some of these may yield to the SPY approach.

## 10  APPENDIX - Matrices of References

Relative/Absolute Referencing ("Dollaring") is the key to making Blocks fill correctly so that there is the same formula in each cell. It also affects how formulas Copy and Paste (but NOT how they Cut and Paste). A brief summary is included below in a matrix. Any deviation from the possibilities listed below will cause errors when filled. Where alternatives are indicated both alternatives will fill correctly but the result of using Copy and Paste will be different for each.

| Normal Reference relative / absolute requirements | Reference to single cell | Reference to single row Block | Reference to single column Block | Reference to multiple row and column Block |
|---|---|---|---|---|
| **Formula in a single cell (not filled)** | Any reference | Any range | Any range | Any range |
| **Formula filled across a single row** | Absolute column any row ($G5 or $G$5) | Relative column any row (G5 or G$5) | Absolute column any row range ($G5:$G8 or $G$5:$G$8) | Relative column any row range (G5:J8 or G$5:J$8) |
| **Formula filled down a single column** | Any column absolute row (G$5 or $G$5) | Any column absolute row range (G$5:J$5 or $G$5:$J$5) | Any column relative row (G5 or $G5) | Any column relative row range (G5:J5 or $G5:$J5) |
| **Formula filled across and down into multiple columns and rows** | Absolute column absolute row ($G$5) | Relative column absolute row (G$5) | Absolute column relative row ($G5) | Relative column relative row (G5) |

The matrix below shows the Complex Reference version of range references. The SUM() function is used in the examples to generate cumulative values. Some other functions such as COLUMNS() or ROWS() can be used in the same way. Careful use of Complex References can provide alternatives to formulas that have CRIT defects (see example below).

| Complex Reference relative / absolute requirements | Cumulative Sum of single row Block | Cumulative Sum of single column Block | Cumulative Sum of all rows of multiple row and column Block | Cumulative Sum of all columns of multiple row and column Block |
|---|---|---|---|---|
| **Formula filled across a single row** | Complex column any row range SUM($G5:G5) or SUM($G$5:G$5) | Not applicable | Not recommended SUM($G5:G8) or SUM($G$5:G$8) | Not applicable |
| **Formula filled down a single column** | Not applicable | Any column complex row range SUM(G$5:G5) or SUM($G$5:$G5) | Not applicable | Not recommended SUM(G$5:J$5) or SUM($G$5:$J$5) |
| **Formula filled across and down into multiple columns and rows** | Not applicable | Not applicable | Complex column relative row range SUM($G5:G5) | Relative column complex row range SUM(G$5:G5) |

|   | C | D | E | F | G |
|---|---|---|---|---|---|
| 3 | 0 | 20 | 65 | 85 | Cumulative |
| 4 |   | =D3-C3 | =E3-D3 | =F3-E3 | Incremental |
| 5 | 1) Incremental with Seed Value | | | | |
| 6 |   |   |   |   |   |
| 7 | 0 | 20 | 65 | 85 | Cumulative |
| 8 |   | =INDEX($C7:$F | =INDEX($C7:$F7,COLUMNS($C7:E7))-INDEX($C7:$F7,COLUMNS($C7:E7)-1) | =INDEX($C7:$ | Incremental |
| 9 | 2) Incremental with Complex Reference | | | | |
| 10 |   |   |   |   |   |